\documentstyle[eqsecnum,aps,epsfig]{revtex}
\newcommand{\Perp}{{\!\perp}}
\newcommand{\V}[1]{{\bf #1}} % Vector
\newcommand{\Vperp}[1]{\V{#1}_\Perp}
\newcommand{\nablap}{\nabla_\Perp}

\newcommand{\BesselK}{{\cal K}}

\newcommand{\vecv}{\V{v}}
\newcommand{\vecb}{\V{b}}
\newcommand{\vecn}{\V{n}}
\newcommand{\nzero}{\V{n}_0}
\newcommand{\vecdn}{\delta\V{n}}

\newcommand{\vecx}{\V{x}}
\newcommand{\qperp}{\Vperp{q}}

\newcommand{\Ftgb}{F_{\text{tgb}}}
\newcommand{\Ftgbi}{F_{\text{tgb}}^{\text{int}}}

\newcommand{\btgb}{\V{b}_{\text{tgb}}}
\newcommand{\btgbc}{b_{\text{tgb}}}

\newcommand{\fTGBi}{f_{\text{TGB}}^{\text{int}}}

\newcommand{\Fel}{F_{\text{el}}}

\newcommand{\ftot}{f_{\text{tot}}}

\newcommand{\fdisl}{f_{\text{disl}}}
\newcommand{\Fcore}{F_{\text{core}}}
\newcommand{\fch}{f_{\text{ch}}}

\newcommand{\FAi}{F_{\text{Abr}}^{\text{int}}}

\begin{document}
\draft
\twocolumn[\hsize\textwidth\columnwidth\hsize\csname@twocolumnfalse%
\endcsname
\title{Dislocation Geometry in the TGB$_A$ Phase: Linear Theory}
\author{Igor Bluestein, Randall D. Kamien, and T.C. Lubensky}
\address{Department of Physics and Astronomy,
   University of Pennsylvania,\\
   Philadelphia, PA 19104}
\date{\today}
\maketitle
\begin{abstract}
We demonstrate that an arbitrary system of screw dislocations in a
smectic-$A$ liquid crystal may be consistently treated within
harmonic elasticity theory, provided that the angles between
dislocations are sufficiently small. Using this theory, we
calculate the ground state configuration of the TGB$_A$ phase. We
obtain an estimate of the twist-grain-boundary spacing and screw
dislocation spacing in a boundary in terms of the macroscopic
parameters, in reasonable agreement with experimental results.
\end{abstract}
%\pacs{PACS numbers: 61.30.Jf, 61.30.-v, 61.30.Cz, 61.72.Gi,
%61.72.Mm}
]
\section{Introduction}
Condensed matter systems offer a vast stage for the intricate
interplay between order and disorder. A beautiful example of such
interplay is the twist-grain-boundary phase (TGB) of chiral
smectics, which is the liquid-crystalline analog of the Abrikosov
vortex state of type II superconductors
\cite{Lubensky1988,LubenskyBook}. Morphologically, the phase
consists of blocks of pure smectic (which can be either
smectic-$A$ or smectic-$C$) separated by parallel, regularly
spaced twist grain boundaries, where each boundary is formed by a
periodic array of screw dislocations. The direction of
dislocation lines rotates by a constant angle from one grain
boundary to the next. Such a dislocation arrangement causes the
smectic blocks to rotate about the axis perpendicular to the
grain boundaries dragging the nematic director along. Thus the
TGB structure combines the properties of smectics and
cholesterics: the nematic director twists on average as in
cholesterics while the lamellar structure of a smectic is
preserved. In this paper only the TGB$_A$ phase will be
considered. As suggested by its name, the smectic blocks in
TGB$_A$ are smectic-$A$.

\begin{figure}[!hbt]
  \begin{center}
  \epsfig{file=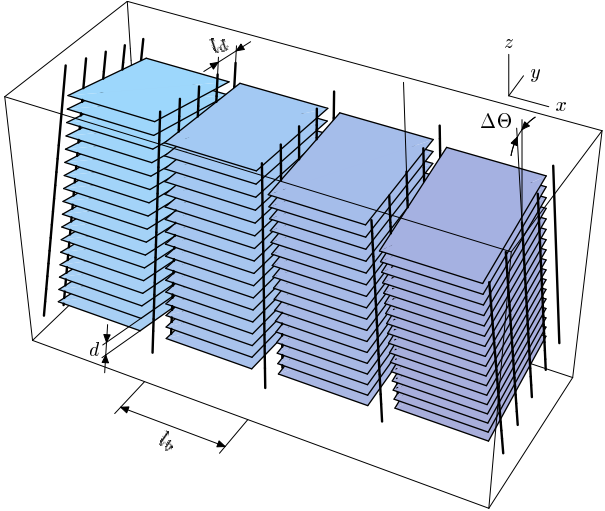,width=80mm}
  \caption{Schematic representation of the TGB$_A$ phase.}
  \label{TGBgraph}
  \end{center}
\end{figure}

The analogy between the TGB$_A$ phase and the Abrikosov vortex
lattice is based on the mathematical similarity of
the Gibbs free energies for metals in a magnetic field and for
chiral
smectics \cite{Lubensky1988,LubenskyBook,deGennes1972,deGennesLC}.
Their respective forms, known as the Ginzburg-Landau free energy
and the de Gennes free energy, are
\begin{eqnarray}
\label{GinzburgLandau} G_{\rm{GL}} = \int &d^3x\bigg[
\displaystyle{\frac{1}{2 m^*}
\left|\left(\frac{\hbar}{i}\nabla-\frac{e^*}{c}\V{A}
\right)\psi\right|^2 +r |\psi|^2}\nonumber\\
&\displaystyle{+\frac{1}{2} g
|\psi|^4 +\frac{1}{8 \pi \mu}(\nabla\!\times\V{A})^2}\nonumber\\
&\displaystyle{-\frac{\V{H}}{4
\pi}\,\cdot\!\nabla\!\times\V{A}}\bigg],
\end{eqnarray}
and
\begin{eqnarray}
\label{deGennes}
  &G_{\rm{deG}} = \displaystyle{\int} d^3 x \big[
     C_\Perp(\nabla-i q_0 \vecn)\psi(\nabla+i q_0\vecn)\psi^*\nonumber\\
      &+(C_\parallel-C_\Perp)n_i n_j
      (\nabla-i q_0 \vecn)_i\psi(\nabla+i q_0\vecn)_j\psi^*\nonumber\\
      &\displaystyle{+r |\psi|^2 + \frac{1}{2} g |\psi|^4
         +\frac{1}{2}K_1(\nabla \!\cdot \vecn)^2}\nonumber\\
    &\displaystyle{+\frac{1}{2}K_2(\vecn\!\cdot \nabla\!\times\vecn)^2
    +\frac{1}{2}K_3\left[\vecn\!\times(\nabla\!\times\vecn)\right]^2}\nonumber\\
    &-h\,\vecn\!\cdot(\nabla\!\times\vecn)\big],
\end{eqnarray}
where $\psi$ is a complex order parameter, $\bf A$ the vector
potential, $\bf H$ the magnetic intensity, $h$ a chiral field
determined by molecular structure \cite{HKL}, and $\vecn$ the unit
director, often decomposed as $\vecn=\vecn_0 +\vecdn$. The
Ginzburg-Landau free energy (\ref{GinzburgLandau}) defines two
characteristic length scales: the order parameter coherence length
$\xi = (\hbar^2/2 m^* |r|)^{1/2}$ and the magnetic field
penetration depth $\lambda = (m^* c^2 g/4 \pi \mu\,e^{*2}
|r|)^{1/2}$. Their ratio $\kappa=\lambda/\xi$, called the Ginzburg
parameter, controls the phase diagram as a function of temperature
and external magnetic field. When the Ginzburg parameter is less
than the critical value $\kappa_c=1/\sqrt{2}$, the system is type
I, and there is a first order transition between a normal metal in
a field and the Meissner phase with $\psi={\rm constant}\ne 0$ and
magnetic field ${\bf B}=0$. When $\kappa>\kappa_c$, the system
becomes type II, and a new phase intervenes between the
normal-metal state and the Meissner state. This new phase, the
Abrikosov flux phase, is characterized by a proliferation of
linear topological defects of the complex order parameter field
$\psi$. The defects are magnetic flux lines, and they form a
two-dimensional triangular lattice throughout the phase.

A very similar phenomenon occurs in chiral smectics, only the
situation becomes somewhat more complicated because of anisotropy
of the de Gennes free energy (we note that the theory of
anisotropic superconductors is equally complex, see \emph{e.g.}
\cite{HaoClemm1992}). Instead of a single order parameter
coherence length $\xi$, there are two lengths: the transverse
coherence length $\xi_\perp = (C_\perp/|r|)^{1/2}$ and the
longitudinal coherence length $\xi_\parallel =
(C_\parallel/|r|)^{1/2}$. But since the values of $C_\perp$ and
$C_\parallel$ can be made equal by rescaling of coordinates, we
will not be concerned with making distinctions between $\xi_\perp$
and $\xi_\parallel$ and assume $\xi_\perp \approx \xi_\parallel
\approx \xi$. The chiral smectic analog of the magnetic field
penetration depth is the twist penetration depth $\lambda_2 = (K_2
g/2 C q_0^2 |r|)^{1/2}$, where $C \approx C_\perp \approx
C_\parallel$.  As in superconductors, the chiral Ginzburg
parameter $\kappa_2 = \lambda_2/\xi$ in liquid crystals determines
the structure of the phase diagram as a function of temperature
and chiral coupling $h$. Again, a defect phase, which is now the
TGB$_A$ phase, appears on the phase diagram when $\kappa_2 >
1/\sqrt{2}$. Linear topological defects in the TGB$_A$ phase are
screw dislocations arranged in twist grain boundaries.

An important problem in the theory of the defect phases predicted
by the Gibbs free energies (\ref{GinzburgLandau}) and
(\ref{deGennes}) is to relate the lattice parameters of defect
lattices to the coupling strengths that enter these energies.
This is rather complicated because the Gibbs energies
(\ref{GinzburgLandau}) and (\ref{deGennes}) include fouth-order
terms that lead to nonlinear Euler-Lagrange equations.
Nevetheless, since the publication of Abrikosov's paper over
forty years ago \cite{Abrikosov1957}, an extensive body of both
experimental and theoretical work on the vortex arrangement in
the Abrikosov phase has been performed, covering all possible
ranges of the external magnetic field, temperature, and Ginzburg
parameter \cite{Fetter1969,Brandt1995}. In contrast with this
happy situation in superconductors, there is only a small
literature that address the liquid-crystal version of the
problem, namely the determination of the grain-boundary spacing
$l_b$ and the dislocation spacing $l_d$ within a grain boundary.
In part this is due to the fact that rotation of defects in
TGB$_A$ makes this version of the problem more complicated.

In their original paper \cite{Lubensky1988} Renn and Lubensky
considered the TGB$_A$ lattice structure near the upper critical
field $h_{c2}$ marking the transition from the cholesteric to the
TGB phase. They employed a model free energy with $C_\perp =
C_\parallel$ and $K_1 = K_3 \equiv K$. Additionally, they assumed
that the twist pitch is very large in comparison to a smectic
layer spacing. Under these assumptions, in a calculation that
parallels Abrikosov's calculation \cite{Abrikosov1957}, they
computed the ratio $l_b/l_d$ for $\kappa_2 = 0.80 > 1/\sqrt{2}$
and various values of $K/K_2$. They found that $l_b/l_d$ increases
from 0.95 for $K/K_2 = 0$ to 1.45 for $K/K_2 = 10^4$. In other
words, $l_b/l_d$ is \emph{sensitive} to the relative values of
the Frank elastic constants. Our approach, which applies near
$H_{c1}$, is applicable when the rotation angle between
consecutive smectic blocks is small, predicts a value close to
0.95 for \emph{all} values of $K/K_2$.

Experimental determination of the TGB$_A$ lattice parameters has
proven to be a difficult task. Early experiments
\cite{Goodby1,Goodby2,Srajer1990,Ihn1992} were crucial in
establishing the existence of the defect phase of chiral smectics
and confirmed that the morphology of this phase agrees with the
predictions of Renn and Lubensky.  However, in all these
experiments $l_b$ and $l_d$ were estimated rather than measured.
Recently an extensive structural study of the TGB$_A$ phase
occurring in a series of chiral tolane derivatives
\cite{Nguyen1992} was undertaken by Navailles and coworkers
\cite{Navailles1998}. They observed an X-ray diffraction pattern
in the transverse direction to the pitch axis that consisted of
discrete Bragg spots, which provided a direct way to measure the
rotation angle $\Delta \Theta$ of smectic blocks. Since the
smectic layer spacing $d$ is easy to measure in the same
experiment, it has now become possible to obtain the precise
values of $l_d$ via $d/2 l_d = \sin(\Delta\Theta/2)$. As an
additional benefit, the information regarding the number of
smectic blocks per pitch provided a way to compute the block size
from the pitch values measured in a different set of experiments
based on the observation of the Grandjean-Cano steps
\cite{Isaert1994}. The ratio $l_b/l_d$ was found to vary from 0.74
to 1.08. Whereas the exact value $l_b/l_d$ is up to
interpretation, these experiments are a clear indication that
this ratio remains close to 1.

In this paper the equilibrium dislocation arrangement in the
TGB$_A$ phase is considered in the limit of low-angle grain
boundaries. Formally, this limit is appropriate close to the
lower critical field $h_{c1}$, marking the transition from the
smectic-$A$ phase to the TGB$_A$ phase. In this case the
dislocation density is low, and it is possible to neglect the
interaction between dislocation cores, so the dislocation core
energy associated with destruction of the order parameter
produces an extensive term in the total free energy. Thus, in
this case, the lattice structure is determined completely by the
elastic energy cost of the distorted smectic structure induced by
the presence of screw dislocations. In the language of
superconductivity, this is the London limit of the Ginzburg-Landau
equations. For our analysis, this means that the results for the
structure of a single vortex and the systems of parallel vortices
obtained in the theory of superconductors may be directly
translated to the language of liquid crystals using the Gibbs
energies (\ref{GinzburgLandau}) and (\ref{deGennes}) as a
dictionary. We will show that there is a simple way to generalize
the results for parallel dislocations to the case of interacting
twist grain boundaries. In the low-angle limit, the rotation
angle is a natural small paramater, and to the lowest order in
this parameter, the interaction energy of two twist grain
boundaries composed of \emph{discrete} linear defects turns out
to be the same as that of grain boundaries with \emph{continuous}
defect distribution. This observation allows us to compute the
energy cost the dislocation arrangement in the TGB$_A$ phase
analytically.

The paper is organized as follows. Section~\ref{elasticity}
introduces the elastic fields suitable for type II smectics-A. We
construct the elastic free energy functional in the harmonic
approximation. We demonstrate that if we have a solution for the
elastic fields uniform in the direction of the layer normal at
infinity, it is possible to construct more general solutions by
superimposing shifted and rotated copies of the original solution.
In Section~\ref{dislocations} we show that in the harmonic
approximation, systems of pure screw dislocations in smectics are
distinguished from other dislocation systems by a constraint that
sets the smectic layers to be minimal surfaces. We compute the
distortion fields and the elastic energy of systems of parallel
screw dislocations.
%The results ofSection \ref{elasticity} are applied to generalize
%the elastic energy calculation for a single screw dislocation to
%arbitrary systems of screw dislocations.
In Section \ref{grain} we calculate the elastic energy of an
individual twist grain boundary and the interaction energy of two
twist grain boundaries. In Section \ref{TGB} we obtain the
elastic energy density for the TGB$_A$ phase. Adding two
extensive terms to it, we construct the total free energy
density. Its minimization yields preferred values for the grain
boundary spacing and the dislocation spacing within a grain
boundary for the given values of material parameters. We find
that in the low-angle regime the ratio $l_d/l_b$ is practically
constant over a wide range of the control parameters, and its
value is 0.95.

\section{Superposition of solutions for distortion fields}
\label{elasticity}
The elastic free energy for smectics-A directly follows from the
de Gennes free energy (\ref{deGennes}) by assuming $|\psi|^2$ is
fixed. Because the smectic density is rapidly varying, it is
standard to write down the elastic free energy of smectics in
terms of the layer displacement field $u$ related to the phase
$\Phi$ of the order parameter $\psi = e^{i \Phi}$ by a
decomposition
\begin{equation}\label{phase}
  \phi \equiv \Phi/q_0 = \nzero\cdot\V{x} - u(\V{x}),
\end{equation}
where $\nzero$ is a fixed unit vector. It is important to note
that in principle $\nzero$ can be chosen to point in an arbitrary
direction and the elastic free energy must be invariant under
rotations of $\nzero$. In some sense, picking $\nzero$ is like
fixing a gauge. Dropping the constant terms, choosing $\nzero$ to
be in the $\hat{z}$ direction, and denoting $\V{n}-\nzero$ as
$\vecdn$, $2 q_0^2 |\psi|^2 C_\parallel$ as $B$, $2 q_0^2 |\psi|^2
C_\perp$ as $D$, we have to quadratic order in $\vecdn$
\begin{eqnarray}\label{fsm}
  F = \frac{1}{2} \int d^3 x&\bigg[B(\partial_z u)^2
   +D(\nablap u + \vecdn)^2
  \nonumber\\&+K_1(\nablap \!\cdot \vecdn)^2+
      K_2(\nablap \!\times \vecdn)^2
   \nonumber\\&+K_3(\partial_z \vecdn)^2\bigg].
\end{eqnarray}
This form of the smectic elastic free energy is known as the
harmonic approximation. Note that the de Gennes free energy and,
consequently, the elastic free energy (\ref{fsm}) derived from it
are invariant with respect to small rotations but not with respect
to arbitrary rotations.  Therefore, their
validity is limited to only those smectic configurations where the
smectic layer normal $\V{N}$ and the nematic director $\V{n}$ do
not deviate significantly from $\nzero$.

The Euler-Lagrange equations derived from the free energy
(\ref{fsm}) are linear, so it is easy to construct new solutions
by superimposing ones already known. To do this, we will
exploit an underlying symmetry of the full theory (\ref{deGennes}).
Under an arbitrary rotation ${\bf R}$, $\phi$ and $\vecn$ transform
as
\begin{eqnarray}
\label{rotPhi}
 &\phi'({\bf x}) = \phi({\bf Rx}),\\
\label{rotN}
 &\vecn'({\bf x}) = {\bf R}^{-1}\vecn({\bf Rx}).
\end{eqnarray}
The displacement field $u'$ inherits its transformation properties
through its definition $u\equiv \vecn_0\cdot\vecx - \phi$. In
general this transformation is nonlinear in the rotation angle
$\theta$. However, since we are interested in small rotations, we
may expand the transformation laws (\ref{rotPhi}), (\ref{rotN}) in
the rotation angle  $\theta$ and keep only the linear terms.

Consider the distortion fields $u$ and $\vecdn$ produced by a
linear source perpendicular to undistorted smectic layers at
infinity. We will construct rotated solutions by modifying simpler
solutions: without loss of generality, we initially take our
source to point along the $\hat{z}$-axis. If we choose $\nzero =
\hat{z}$, the distortion fields will be independent of $z$:
\begin{eqnarray}
  &\phi(x,y,z) = z - u(x,y),\\
  &\vecdn(x,y,z) = \vecdn(x,y).
\end{eqnarray}
We may construct a solution which corresponds to a superposition
of sources that intersect the $xy$-plane at $(x^\alpha,y^\alpha)$
and oriented along $\V{N}_0^\alpha = \hat{z}+ \theta^\alpha
\V{t}^\alpha$, where $\V{t}^\alpha$ is a unit vector orthogonal to
$\hat{z}$ and $\theta^\alpha \ll 1$. First, consider what happens
to the original solution if the system undergoes a rigid physical
rotation about $x$-axis by a small angle $\theta$. The phase
function $\phi$ is a scalar, so the transformed function is just
the original function of the transformed coordinates:
\begin{equation}
    \phi'(x,y,z) = \phi(x',y',z')
    = \phi(x,y-\theta z,z+\theta y).
\end{equation}
In terms of the original elastic field $u$, the transformed phase
function is
\begin{equation}\label{phiPrime}
  \phi'(x,y,z) = z+\theta y - u(x,y-\theta z).
\end{equation}
We can read the decomposition (\ref{phiPrime}) in two ways. In the
first version, it defines the rotated elastic field $u'(x,y,z)$
with respect to $\nzero' = \hat{z}+\theta\hat{y}$:
 \begin{equation}\label{uPrime}
  u'(x,y,z) = u(x,y-\theta z).
\end{equation}
The corresponding transformation of $\vecdn$ is
\begin{equation}\label{dnPrime}
  \vecdn'(x,y,z) = \vecdn(x,y-\theta z).
\end{equation}
If we consider the effect of rotations in Fourier space, we note
that if
\begin{equation}
   g(x,y,z)= \int \frac{d^3 q}{(2 \pi)^3}\,g(q_x,q_y,q_z)
   e^{i q_x x} e^{i q_y y} e^{i q_z z},
\end{equation}
then it follows that
\begin{equation}\label{rotateFourier}
   g'(q_x,q_y,q_z) \approx g(q_x,q_y-\theta q_z,q_z+\theta q_y).
\end{equation}

The second interpretation of (\ref{phiPrime}) is as a definition
of the rotated field $u''$ with respect to the original
$\nzero=\hat{z}$. In this case, the transformed elastic fields
acquire an extra asymptotic term, linear in $y$:
\begin{eqnarray}
\label{uDoublePrime}
   u''(x,y,z) &= -\theta y + u(x,y-\theta z),\\
\label{dnDoublePrime}
   \vecdn''(x,y,z) &= \theta \hat{y} + \vecdn(x,y-\theta z).
\end{eqnarray}
Even if we have a superposition of sources of different
orientation, the elastic fields $u_\alpha''$, $\vecdn_\alpha''$
produced by individual sources are all defined with respect to the
same fiducial choice of $\nzero = \hat{z}$, so it is consistent to
superimpose these fields and claim that this superposition of
fields is the solution for the superposition of sources:
\begin{eqnarray}
\label{super1}
  &u_{\rm super}(x,y,z)&=\\
    &&\sum_\alpha[\,-\theta^\alpha\,\V{t}^\alpha\!\cdot\V{x}
    + u(\Vperp{x}-\Vperp{x}^\alpha-\theta^\alpha \V{t}^\alpha z)],\nonumber\\
\label{super2}
  &\vecdn_{\rm super}(x,y,z) &=\\
    &&\sum_\alpha [\,\theta^\alpha \V{t}^\alpha
    + \vecdn(\Vperp{x}-\Vperp{x}^\alpha-\theta^\alpha \V{t}^\alpha z)].\nonumber
\end{eqnarray}
where $u_{\rm super}$ and $\vecdn_{\rm super}$
are the total displacement and director
fields, respectively.
The asymptotic parts of the distortion fields are absolutely
essential in formulating global geometric constraints on smectic
configurations, in particular, for distinguishing different types
of dislocations.  When constructing linear superpositions of defects we
must make sure that the layer normals are unambiguously
defined everywhere.  Fortuitously, the terms linear in $y$ drop out
of the {\sl energetics}.  As a result we are free
to use either $u'(x,y,z)$ or $u''(x,y,z)$ in our calculations.
We will
exploit this fact in the next section.

\section{Interaction energy of screw dislocations}
\label{dislocations}
\subsection{Screw dislocations vs. edge dislocations}

Dislocations in smectics are linear topological defects of the
lamellar structure \cite{LubenskyBook,KlemanBook}. In the presence
of dislocations, it becomes impossible to devise a consistent
global numbering scheme for the smectic layers. In other words,
the phase field $\Phi$ cannot be defined as a continuous field
for the whole region where the smectic order parameter is
defined. If a closed contour surrounds a dislocation core and one
insists that $\Phi$ be continuous, then the phase differences
along the contour will add up to a multiple of $2 \pi$:
\begin{equation}\label{intPhi}
  \oint d \Phi = 2 \pi n.
\end{equation}
However, it is possible to break the space into a number of
overlapping regions and define $\Phi$ for each region in such
a way that local definitions of $\Phi$ differ in the
intersections of regions only by a constant. In this case the
gradient field $\nabla \Phi$ is globally defined (single-valued
and continuous). If we
choose the same ``gauge'' $\nzero$ for each region, then the
field $\vecv\equiv -\nablap u$ defined by $\nabla \phi \equiv \nabla (\Phi/q_0)
= \nzero+\vecv$ will be also globally consistent. In terms of
$\vecv$, the above integral can be rewritten as
\begin{equation}\label{intV}
  \oint \vecv\cdot d\V{l} = n d.
\end{equation}
It is convenient to represent a dislocation line of strength $n$
and position $\vecx = \vecx(l)$ by a singular dislocation density
\begin{equation}\label{dSource1}
  \vecb(\vecx) = d \int d l\, \frac{d \vecx(l)}{d
  l}\,n\,\delta^{(3)}\!\!\left(\vecx-\vecx(l)\right).
\end{equation}
With its help, the integral relation (\ref{intV}) can be turned
into a differential form:
\begin{equation}\label{curlV0}
  \nabla \!\times \vecv = \vecb(\vecx).
\end{equation}
For a system of dislocations, the density is constructed by
superimposing densities of individual lines in the form
(\ref{dSource1}). Clearly, this procedure guarantees that all
integrals in the form (\ref{intPhi}) or (\ref{intV}) will have the
right value. In the most general case, distributions of
dislocations can be continuous. In principle, any function
$\vecb(\vecx)$ can be considered to be a dislocation density as
long as it satisfies a conservation law $\nabla\!\cdot\V{b} = 0$,
which says that dislocation lines cannot end inside the system.

The integral constraints (\ref{intPhi}) and (\ref{intV}) or their
differential equivalent (\ref{curlV0}), define dislocations
topologically, but by their nature they are incapable of
determining how dislocations are actually arranged in physical
systems. To understand that, we need to look at dislocations as
physical objects that have energy. There are two contributions to
the dislocation energy. The core energy $\Fcore$ arises from the
destruction of the order parameter in the core region. It is
proportional to the total length of dislocation lines in the
system, but does not depend on the details of the smectic
configuration outside the core. Still, for topological reasons,
the smectic structure outside the core is necessarily distorted,
and these distortions give rise to the elastic energy $\Fel$ of
the dislocation.

The structure of the distortion fields induced by a dislocation
depends significantly on the orientation of the dislocation core
with respect to the smectic layers. This orientation reflects the
physical nature of the dislocation. We may consider the Volterra
construction for two classes of defects \cite{LubenskyBook}. One
way to produce a dislocation in a smectic is to force the layers
to wind in a helical fashion, which causes the core to be
perpendicular to the smectic layers at infinity. A physically
different procedure is to remove half a layer, which makes the
core parallel to the layers. Dislocations of the first type are
known as screw dislocations and of the second type as edge
dislocations.

Now let us turn to calculation of the elastic energy of a system
of screw dislocations. The program is to construct an appropriate
dislocation density $\vecb(\vecx)$ and then minimize the smectic
elastic free energy while requiring compliance with
(\ref{curlV0}). If we restrict our consideration to systems of
nearly parallel screw dislocations with separations
large compared to the preferred layer
spacing $d$, then the criteria for the validity of the harmonic
approximation are met and we can use the elastic free energy
(\ref{fsm}).

First, consider a single screw dislocation at the origin. The
corresponding dislocation density is
\begin{equation}\label{b1}
  \V{b}_1(\V{x})= \hat{z}\,d\,\delta(x)\delta(y).
\end{equation}
If we make the natural choice $\nzero = \hat{z}$, the distortion fields
$\vecv$ and $\vecdn$ are cylindrically symmetric. It turns out
that the cylindrical symmetry and the topological condition
(\ref{curlV0}) completely determine the $\vecv$-field.

An immediate consequence of the cylindrical symmetry of $\vecv$
and $\vecdn$ is that their derivatives in the $z$-direction
vanish: $\partial_z \vecv = 0$, $\partial_z \vecdn = 0$. The $x$-
and $y$-components of the topological condition (\ref{curlV0}),
imply that $\partial_y v_z =0$ and $\partial_x v_z =0$. It then
follows that $v_z$ is constant, set to zero by the boundary
conditions at infinity:
\begin{equation}\label{vz}
  v_z = 0.
\end{equation}
Therefore, the layer displacement field $u$ is independent of $z$,
and the distortion fields induced by a single screw dislocation
are of the type considered in the previous section. Thus, we may
construct the distortion fields for the entire system of nearly
parallel screw dislocations by the superposition procedure of
(\ref{super1}) and (\ref{super2}).

There is another constraint on the distortion fields imposed by
the cylindrical symmetry. Since $\partial_z \vecv$ and $\partial_z
\vecdn$ vanish, the compression term and the bend term drop out
from the elastic free energy~(\ref{fsm}). The remaining terms
involve only the radial and azimuthal components of the distortion
fields. Moreover, there is no term that mixes different
components. Observe that the topological condition~(\ref{curlV0})
fixes the azimuthal component of $\vecv$. Among all configurations
of the distortion fields with fixed azimuthal components, the
configuration with $v_\rho = 0$, $\delta n_\rho = 0$ has the
lowest energy. This means that the distortion fields induced by a
single screw dislocation are not only cylindrically symmetric, but
also have a single azimuthal component. A vector field of such
structure is solenoidal, and we can conclude that
\begin{eqnarray}
\label{divV0}
   \nabla\!\cdot\vecv = 0, \\
\label{divDn}
   \nabla\!\cdot\vecdn = 0.
\end{eqnarray}
The last relation eliminates the splay contribution to the elastic
free energy. Although (\ref{vz}), (\ref{divV0}), and (\ref{divDn})
are obtained for a single screw dislocation, the superposition
procedure guarantees that they remain valid even for the
distortion fields produced by arbitrary systems of screw
dislocations as long as the harmonic approximation is applicable.

A simple argument shows that (\ref{divV0}) cannot hold in the
presence of edge dislocations. Consider a single edge dislocation
oriented in the $y$-direction.
As in the case of a screw dislocation, the translational symmetry
of the distortion fields in the core direction together with the
topological condition (\ref{curlV0}) makes the $\vecv$-field
orthogonal to the core direction, so $v_y = 0$. Then
$\nabla\cdot\vecv$ reduces to $\partial_x v_x+\partial_z v_z$. It
is easy to check that $\partial_x v_x$ and $\partial_z v_z$ have
the same sign everywhere, so $\nabla\cdot\vecv$ never vanishes.
The presence of additional edge dislocations does not modify this
conclusion.

Thus we see that in the harmonic approximation, the constraint
(\ref{divV0}) provides a very clear distinction between systems
of pure screw dislocations and other dislocation systems. Note
that (\ref{vz}) and (\ref{divV0}) imply that smectic layers in
the presence of screw dislocations are minimal surfaces, which
are defined as surfaces of zero mean curvature. If $\V{N}$ is a
field of unit normals to a family of surfaces $\phi(x,y,z) =
\textrm{const}$ filling the space, then the mean curvature of the
surfaces in the family is proportional to
$\nabla\!\cdot\V{N}$~\cite{FrankelBook}, the multiplicative
constant chosen by convention. The field of unit normals to the
smectic layers is
\begin{equation}\label{minCurv}
  \V{N} = \frac{\nabla \phi}{|\nabla \phi|}
    = \frac{\hat{z}+\vecv}{\sqrt{1+2 v_z + \vecv^2}}.
\end{equation}
Since $v_z$ vanishes, $|\nabla \phi| = 1+O(\vecv^2)$. So in the
harmonic approximation the denominator is unity, and
\begin{equation}
  \nabla\!\cdot\V{N} = \nabla\!\cdot\vecv = 0.
\end{equation}

\subsection{Elastic energy of screw dislocations}

Following \cite{LubenskyBook}, we can obtain an explicit solution
for the distortion fields induced a single screw dislocation. The
$\vecv$-field is determined by equations (\ref{curlV0}) and
(\ref{divV0}) with $\V{b}(\V{x})$ given by (\ref{b1}). At infinity
the smectic layers are undistorted, so $\vecv \to 0$ as $\rho \to
\infty$. To emphasize that the problem is in fact two-dimensional,
we rewrite (\ref{curlV0}) and (\ref{divV0}) as
\begin{eqnarray}
\label{curlV}
  \nablap \!\times \vecv &= \vecb(\vecx), \\
\label{divV}
  \nablap \!\cdot \vecv &= 0.
\end{eqnarray}
For our purposes, it is convenient to solve these equations in
Fourier space. A general solution to (\ref{curlV}) can be
decomposed into longitudinal and transverse components with
respect to the direction of $\qperp$. Equation~(\ref{divV})
eliminates the longitudinal component, so we have
\begin{equation}\label{solnLayers}
  \vecv(\qperp) =
   i \frac{\qperp\!\times \V{b(\qperp)}}{\qperp^2},
\end{equation}
where in the case of a single screw dislocation $\V{b({\bf q})} =
\hat{z} d\delta(q_z)$. The nematic director tilt field $\vecdn$ can be
obtained from the Euler-Lagrange equations derived from the
elastic free energy (\ref{fsm}). Earlier we discovered that the
compression, bend and splay terms drop out. Then the elastic free
energy reduces to
\begin{eqnarray} \label{freeEnergy}
  F &= \frac{1}{2} \displaystyle\int d^3 x \left[D(\nablap u + \vecdn)^2
       + K_2(\nablap \!\times \vecdn)^2\right]\nonumber\\
    &= \frac{1}{2}
       \displaystyle\int \frac{\displaystyle d^3 q}{\displaystyle(2 \pi)^2}\,
       \Bigg[\,D|\vecv(\qperp)-\vecdn(\qperp)|^2\nonumber\\
       &+K_2|\qperp\!\times\vecdn(\qperp)|^2\,\Bigg].
\end{eqnarray}
Variation of (\ref{freeEnergy}) with respect to $u$ and $\vecdn$
leads to the following Euler-Lagrange equations:
\begin{eqnarray}
\label{euler1}
  &\displaystyle{\nablap\!\cdot(\vecv-\vecdn) = 0},\\
\label{euler2}
  &\displaystyle{\nablap^2 \vecdn = \frac{1}{\lambda^2}\left(\vecdn-\vecv\right)},
\end{eqnarray}
where $\lambda \equiv \lambda_2 = (K_2/D)^{1/2}$ is the twist
penetration depth. The first equation is automatically satisfied.
The solution of the second equation in Fourier space is
\begin{equation}\label{solnTilt}
  \vecdn(\qperp)
   = \frac{1/\lambda^2}{\qperp^2+1/\lambda^{2}}\,\vecv(\qperp).
\end{equation}
In real space, the solutions for the distortion fields in polar
co\"ordinates $\rho$ and $\phi$ are
\begin{eqnarray}
\label{solnLayers1}
  &\vecv = \displaystyle{\frac{d}{2 \pi \rho}}\,\V{e}_\phi,\\
\label{solnTilt1}
  &\vecv-\vecdn = \displaystyle{\frac{d}{2 \pi \lambda}}\,
      \BesselK_1(\rho/\lambda)\,\V{e}_\phi,
\end{eqnarray}
where $\BesselK_1$ is the modified Bessel function of order $1$.
Substituting the distortion fields (\ref{solnLayers}) and
(\ref{solnTilt}) into the elastic free energy (\ref{freeEnergy}),
we find that the elastic energy cost per unit length of a single
screw dislocation is
\begin{equation}\label{fOneDisl}
 \frac{\Fel^{(1)}}{L}
   = \frac{1}{2} D d^2 \int \frac{d^2 q}{(2 \pi)^2}\,
     \frac{1}{\qperp^2+1/\lambda^{2}}.
\end{equation}
Since the smectic order exists only outside the dislocation core,
the integration region has to be restricted to the disk $|\qperp|
< 1/a$, where $a$ is the core radius. Then the integral gives
\begin{equation}
   \frac{\Fel^{(1)}}{L}
     = \frac{D d^2}{8 \pi}\,\ln \left(\frac{\lambda^2}{a^2}+1\right).
\end{equation}
Note that in the extreme type I systems $\lambda \to 0$, so there
is no elastic contribution to the screw dislocation energy in the
harmonic approximation \cite{deGennesLC,KlemanBook}.

It is straightforward to generalize (\ref{fOneDisl}) to an arbitrary
density of screw dislocations parallel to the $z$-axis:
\begin{equation}\label{manydisl}
 \frac{\Fel}{L} = \frac{1}{2} D \int \frac{d^2 q}{(2 \pi)^2}\,
     \frac{1}{\qperp^2+1/\lambda^{2}}\vert \bar b(\qperp)\vert^2,
\end{equation}
where the scalar areal dislocation density $\bar{b}(\qperp)$ is
defined via $\V{b}({\bf q}) \equiv \hat{z} \bar{b}(\qperp)\delta(q_z) =
\hat{z} d \delta(q_z)\sum_\alpha \exp\{i q_x x_\alpha\} \exp\{i q_y y_\alpha\}$.

\section{Interaction of twist grain boundaries}
\label{grain} A twist grain boundary separates two smectic domains
with layer normals that, while pointing in different directions,
remain perpendicular to some axis. Physically, it can be
implemented as an array of equidistant parallel screw
dislocations. In this case topology imposes a constraint on the
layer rotation angle $\Delta \Theta$. Rewriting the topological
condition (\ref{intPhi}) in terms of the layer normal and the
local layer spacing $d(\V{x})$ \cite{deGennesLC},
\begin{equation}
  \oint \frac{\V{N}}{d(\V{x})}\cdot d\V{l} = n,
\end{equation}
If we consider a rectangular integration path in the $xy$-plane
which surrounds one
$n=1$ defect then, with $d$ fixed, we find that
\begin{equation}
  \frac{l_d}{d} \left[\delta{\bf N}_+ - \delta{\bf N}_-\right] =1,
\end{equation}
where $\delta{\bf N}$ is the projection of $\bf N$ onto the
$xy$-plane, $d$ is the equilibrium layer spacing and $l_d$ is the
dislocation spacing. This change corresponds to the rotation of
smectic layers by $\Delta \Theta = 2 \sin^{-1}(d/2 l_d)$, which
becomes $d/l_d$ in the low-angle limit. It should be emphasized
that although the rotation angle of the smectic layers is dictated
by topology, topology in no way requires a specific orientation of
defects in the grain boundary with respect to the smectic layers
at infinity. Rather, this orientation is determined energetically.
Leaving a detailed comparison of energetics of various dislocation
systems outside the scope of this paper, here we will assume that
pure screw dislocation systems are energetically preferrable to
systems of screw-edge dislocations with similar geometry and
consider only those configurations of defects and smectic layers
that correspond to pure screw dislocation systems. The formalism
developed in the preceding section applies specifically to this
kind of system.

The above result regarding the rotation angle of the
smectic layers can be also obtained by considering the distortion
fields. If the plane of the boundary is the $yz$-plane and the
dislocations are parallel to the $z$-axis, then the dislocation
source has the following form:
\begin{equation}\label{oneWall}
  \btgb(x,y,z) = \hat{z} d\delta(x) \sum_{n = -\infty}^{\infty} \delta(y-nl_d),
\end{equation}
where $l_d$ is the spacing between defects along the $y$-axis.
The layer tilt $\vecv$ induced by the source (\ref{oneWall}) can
obtained by superimposing the contributions of individual
dislocations (\ref{solnLayers1}):
\begin{eqnarray}
  v_x(x,y) &= \displaystyle{\frac{d}{2 \pi}
    \sum_{n = -\infty}^\infty
    \frac{\displaystyle y - n l_d}{\displaystyle x^2+(y-n l_d)^2}}, \\
  v_y(x,y) &= -\displaystyle{\frac{d}{2 \pi}
  \sum_{n = -\infty}^\infty}
    \frac{x}{x^2+(y-n l_d)^2}.
\end{eqnarray}
The sums can be computed explicitly with the help of the Poisson
summation formula \cite{Kamien1999}:
\begin{eqnarray}
\label{v1x}
  v_x &= \displaystyle{\frac{d}{2 l_d} \frac{\sin 2 \pi y/l_d}
   {\cosh 2 \pi x/l_d - \cos 2 \pi y/l_d}},\\
\label{v1y}
  v_y &= -\displaystyle{\frac{d}{2 l_d} \frac{\sinh 2 \pi x/l_d}
   {\cosh 2 \pi x/l_d - \cos 2 \pi y/l_d}}.
\end{eqnarray}
The limiting form of $v_x$ and $v_y$ for large $|x|$ is
\begin{equation}
  \vecv(x\to\pm \infty) = \pm \frac{d}{2 l_d}\,\hat{y},
\end{equation}
which shows that the smectic layers undergo a rotation by $d/l_d$
about the $x$-axis as they cross the dislocation array at $x=0$.
While the director relaxes to the layer normal in a distance of the
order of the twist penetration depth $\lambda$, we see that
the smectic layers relax to the undistorted asymptotic configuration
within a distance $l_d$ of the grain boundary.

The intensive energetic characteristic of a twist grain boundary is
the elastic energy per unit area. It can be computed by the
following limiting procedure. Instead of an infinite dislocation
array (\ref{oneWall}), consider an array of $N$ screw
dislocations. The array extension in the $y$-direction is $N
l_d$. Its elastic energy $\Fel(N)$ is given by (\ref{manydisl}).
Then the elastic energy per unit interval of the $y$-axis of a
complete twist grain boundary can be taken as a limit of the
ratio of the energy of a finite array to its extension in the
$y$-direction:
\begin{equation}\label{infLimit1}
  \frac{\Ftgb^{(1)}}{l_d} = \lim_{N\to\infty}\frac{\Fel(N)}{N l_d}.
\end{equation}
To implement this limit,  we should consider
the square of the amplitude of the areal dislocation density
\begin{equation}
   |\bar b|^2 = d^2 \sum_{n=1}^N\sum_{n'=1}^N
       e^{iq_yl_d (n-n')}.
\end{equation}
As $N\rightarrow\infty$, note that $|\bar b|^2/N \rightarrow \sum_{n=-\infty}^\infty
\exp\{iq_yl_dn\}$ and thus
the energy per unit area of a twist grain boundary is
\begin{equation}\label{Ftgb1}
  \frac{\Ftgb^{(1)}}{A}
    = \frac{D d^2}{2 l_d} \sum_{n = -\infty}^\infty
      \int \frac{d^2 q}{(2 \pi)^2}\,
      \frac{e^{i q_y l_d n}}{\qperp^2+1/\lambda^2}.
\end{equation}
Note that $\Ftgb/A$ may be broken into an
extensive part and the interaction part. This corresponds to a
separation of the $n = 0$ term from the rest of the sum:
\begin{equation}
  \frac{\Ftgb^{(1)}}{A} = \frac{1}{l_d}\frac{\Fel^{(1)}}{L}+\frac{\Ftgbi(1)}{A}.
\end{equation}
The interaction part $\Ftgbi$ can be written as a sum over contributions of
dislocation pairs at distances $l_d$, $2 l_d$, \ldots :
\begin{eqnarray}\label{Ftgbi1}
    \frac{\Ftgbi(1)}{A} &= \displaystyle{\frac{D d^2}{2 l_d}
    \sum_{{n = -\infty}\atop {n \neq 0}}^\infty
    \int \frac{d^2 q}{(2 \pi)^2}\,
    \frac{e^{i q_y l_d n}}{\qperp^2+1/\lambda^2}} \nonumber\\
  &= \displaystyle{\frac{D d^2}{2 l_d}
    \sum_{n = 1}^\infty
    \int_{-\infty}^\infty \frac{d q_y}{2 \pi}\,
    \frac{\cos q_y l_d n}{\sqrt{q_y^2+1/\lambda^2}}}\nonumber\\
   &= \displaystyle{\frac{D d^2}{2 \pi l_d}
      \sum_{n = 1}^\infty \BesselK_0\left(\frac{l_d n}{\lambda}\right)},
\end{eqnarray}
where $\BesselK_0$ is the modified Bessel function of order $0$. As
expected, this result closely resembles the interaction energy of
parallel vortices in the London limit \cite{deGennesSC,Tinkham}:
\begin{equation}
    \frac{\FAi}{A}= \frac{1}{n_L}\,\frac{\Phi_0^2}{8 \pi^2 \lambda^2}
    \sum_i \sum_{j > i} \BesselK_0\left(\frac{r_{i
    j}}{\lambda}\right),
\end{equation}
where $\Phi_0 = 2 \pi \hbar c/e^*$ is a quantum of the magnetic
flux, $n_L$ is the vortex areal density, $\lambda$ is the magnetic
field penetration depth.

Our next goal is to compute the interaction energy of twist grain
boundaries. Before we consider the grain boundary system in the
TGB$_A$ phase, let us limit our consideration to systems of finite
number of parallel low-angle grain boundaries that are
sufficiently separated so that the harmonic approximation
(\ref{fsm}) is applicable. Again, topology completely determines
the relative orientations of the smectic layers in different
smectic blocks. To ensure that the defects are pure screw
dislocations, we require that the defects in adjacent grain
boundaries be rotated by $\Delta\Theta = d/l_d$ and, in addition,
that the outermost smectic blocks be rotated by the same angle,
but in the opposite directions with respect to the defects in the
middle of the system. In comparison with the calculation for a
single grain boundary, we are dealing now with more general defect
systems where not all defects are parallel to each other. Our
formalism easily handles this situation.

Consider a system that contains only two twist grain boundaries separated
by a
distance $l$ from each other. The angle between directions of the
dislocation lines in these two boundaries is $\theta = \Delta
\Theta = d/l_d$.  We may implement
this by taking one grain boundary at $x=-l/2$, rotated by
$-\theta/2$ and the other boundary at $x=l/2$, rotated by
$\theta/2$ where $\theta=d/l_d$.  The dislocation density for this
complexion is the sum of the contributions from the two grain
boundaries, $\V{b} = \btgb^{(1)} + \btgb^{(2)}$, where
\begin{eqnarray}
   &\btgb^{(1)}(q_x,q_y,q_z)= \left[{\hat z} - (\theta/2)\hat y\right]\,
          2\pi d\,e^{-iq_xl/2}\times  \nonumber\\
   &\qquad\delta[q_z-(\theta/2) q_y]
   \displaystyle{\sum_{n=-\infty}^\infty}
   e^{i(q_y+\theta q_z/2)nl_d},\\
   &\btgb^{(2)}(q_x,q_y,q_z)= \left[{\hat z} + (\theta/2)\hat y\right]\,2\pi
   d\,e^{iq_xl/2}\times \nonumber\\
    &\qquad\delta[q_z+(\theta/2)q_y]
    \displaystyle{\sum_{n=-\infty}^\infty}
   e^{i(q_y -\theta q_z/2) nl_d}.
\end{eqnarray}
We may transform the sums in the above expression into a sum of
delta functions via the Poisson summation formula:
\begin{eqnarray}
\label{twoboundaries}
   &\btgb^{(1)}(q_x,q_y,q_z) = \left[{\hat z} - (\theta/2)\hat y\right]
       \delta[q_z-(\theta/2)q_y] \times \nonumber \\
   &\qquad\displaystyle{\frac{2\pi d}{l_d} e^{-iq_xl/2}
   \sum_{m=-\infty}^\infty \delta\left[q_y+\theta q_z/2 - \frac{2\pi
   m}{l_d}\right]},\\
\label{twoboundariesa}
   &\btgb^{(2)}(q_x,q_y,q_z)=\left[{\hat z} +(\theta/2)\hat y\right]
         \delta[q_z+(\theta/2)q_y]\times\nonumber \\
   &\qquad\displaystyle{\frac{2\pi d}{l_d}e^{iq_xl/2}
   \sum_{m=-\infty}^\infty \delta\left[q_y -\theta q_z/2-\frac{2\pi
   m}{l_d}\right]}.
\end{eqnarray}
By virtue of linear superposition, the energy of this dislocation
density will be the sum of three terms.  Two of these terms are
simply the self-energies of the two individual grain boundaries
which we have calculated above.  The interaction energy comes from
the cross term which is of the form:
\begin{equation}
\Ftgbi(2,l) = 2\int \frac{d^3q}{(2\pi)^3}\,\btgbc^{(1),i}({\bf
q})M_{ij}({\bf q})\btgbc^{(2),j}(-{\bf q}),
\end{equation}
where $M_{ij}({\bf q})$ is the general interaction kernel that
accounts for both screw and edge dislocations \cite{LubenskyBook}.
Upon inspection of (\ref{twoboundaries}) and
(\ref{twoboundariesa}), we note that: 1) Since $(d/l_d)=\theta$
the $\hat y$-components of the dislocation densities may be
ignored since we are only working to quadratic order in
$\theta=\Delta\theta$, and 2) in the product $\btgbc^{(1)}
\btgbc^{(2)}$ only the $m=0$ terms in (\ref{twoboundaries})
contribute because of the delta-function constraints.  As a
result, only the $q_z=0$ and $q_y=0$ modes of $\btgb^{(1)}$ and
$\btgb^{(2)}$ contribute. Moreover, since the $q_y=0$ modes of
the layer normals in (\ref{v1x}) and (\ref{v1y}) relax to their
asymptotic values immediately, we see that we have constructed a
globally consistent layer structure between the boundaries. Thus
we may use (\ref{manydisl}) to evaluate the interaction energy:
\begin{eqnarray}\label{Ftgbi2}
   \frac{\Ftgbi(2,l)}{A} &=&\displaystyle{
     \frac{D d^2}{2 l_d}\frac{1}{l_d}
     \int_{-\infty}^\infty \frac{d q_x}
     {2 \pi}\frac{e^{i q_x l}+e^{-i q_x l}}{q_x^2+1/\lambda^2}}\nonumber \\
    &=& \displaystyle{\frac{D d^2}{2 l_d}\frac{\lambda}{l_d}\,e^{-l/\lambda}}.
\end{eqnarray}
Note that since
only the $q_y=0$ modes contribute, the interaction energy
is independent of arbitrary
phason shifts of the grain boundaries along $y$. In the harmonic
approximation, the interaction energy of grain boundaries in a
system of several twist grain boundaries breaks down into the sum
of contributions of individual grain boundary pairs. Since the
harmonic theory admits linear superposition, the
result (\ref{Ftgbi2}) can be applied to any angle of rotation
at any separation $l$ as long as the grain boundaries are
composed purely of screw defects.

Another way to look at this result is to consider the behavior of
the director between the grain boundaries.  We may expand the
director and displacement fields in Fourier modes at all the
reciprocal lattice vectors $\V{G}$. In the harmonic approximation,
director and displacement fields from different sources add
linearly. Thus if no pairs of grain boundaries have dislocation
axes parallel or antiparallel and dislocations in each boundary
are straight, each distortion (displacement
$u_{\scriptscriptstyle\bf G}$ or
$\vecdn_{\scriptscriptstyle\bf G}$)
for a given reciprocal lattice vector $\V{G} \neq
0$ arises from a unique grain boundary. Thus finite reciprocal
lattice vector distortions from different grain boundaries do not
interact. Each grain boundary, however, produces a $\V{G} = 0$
director distortion $\vecdn_0$, which is sensitive to the
presence of other grain boundaries. The origin of interactions
between grain boundaries is thus $\vecdn_0$, and we can calculate
these interactions by applying appropriate boundary condition to
$\vecdn_0$. For an isolated grain boundary, $\vecdn_0=(0,\delta
n_y,0)$ reaches constant asymptotic values of $(0, \pm \theta)$.
If there is more than one grain boundary, $\vecdn_0$ has to
rotate through the angles determined by the dislocation complexion in a
shorter distance and at greater energy cost. Consider two walls
with dislocation separation $l_d$ located at $x = \pm l/2$. If we
consider (\ref{euler1}) and (\ref{euler2}) in this situation we
see that $\nablap\!\cdot\vecv=\nablap\!\cdot\vecdn=0$ since both
$\vecv$ and $\vecdn$ only have components along $\hat y$ but only
depend on $x$.  Thus (\ref{euler2}) becomes:
\begin{equation}\label{this}
   \partial_x^2\delta n_y = \frac{1}{\lambda^2}\left[\delta n_y+\nablap u
      \right].
\end{equation}
We have three regions to consider.  For
$x\le-l/2$ we have
\begin{eqnarray}
 &\nablap u = \theta\hat{y},\\
 &\delta n_y = -\theta + Ae^{(x+l/2)/\lambda},
\end{eqnarray}
while between $x=-l/2$ and $x=l/2$
\begin{eqnarray}\label{inbetween}
  &\nablap u = 0,\\
  &\delta n_y=\displaystyle{B\frac{\sinh(x/\lambda)}{\sinh(l/2\lambda)}},
\end{eqnarray}
and for $x\ge l/2$
\begin{eqnarray}
  &\nablap u = -\theta \hat{y},\\
  &\delta n_y = \theta - Ce^{-(x-l/2)/\lambda}.
\end{eqnarray}
Continuity of the director forces $C=A=\theta -B$.  Inserting
these solutions into
into the energy (\ref{freeEnergy}) we have
\begin{equation}
  \frac{F(l)}{A}=D\lambda\left[(B-\theta)^2 + B^2\coth(l/2\lambda)\right].
\end{equation}
Minimizing over the free parameter $B$ and using $\theta=d/l_d$,
we find
\begin{eqnarray}
  \frac{F(l)}{A} &= \displaystyle{\frac{D\lambda d^2}{l_d^2}
  \frac{\coth(l/2\lambda)}{1+\coth(l/2\lambda)}}\nonumber \\
   &= \displaystyle{\frac{D\lambda d^2}{2l_d^2}\left[1 + e^{-l/\lambda}\right]}.
\end{eqnarray}
The
energy of interaction is simply
\begin{equation}
\displaystyle{\frac{\Delta F}{A} = \frac{F(l)-F(\infty)}{A} =
\frac{Dd^2} {2l_d^2}{\lambda e^{-l/\lambda}}}.
\end{equation}
which agrees, with our previous result (\ref{Ftgbi2}).  This shows that
the energy of interaction comes from the ``confinement'' of the director -- it
is forced to twist from $-\theta/2$ to $\theta/2$ in a length on
the order of a few $l$.
\section{TGB$_A$ phase}
\label{TGB}
In the preceding section we showed that in a system composed of
finite number of low-angle twist grain boundaries the energy of
the dislocation interaction within a grain boundary and the
energy of the inter-boundary interaction decay exponentially with
distance. It suggests that the dislocation arrangement in the
TGB$_A$ phase could be treated as well within the same
computational framework, even though the angles between the
directions of dislocations in the entire system are not
restricted to only small ones. Indeed, when the grain boundaries
are low angle, the dislocations that cannot be described as
nearly parallel are separated by the distance of many grain sizes
$l_b$. It is reasonable to hope that we can find a regime
where $l_b$ is sufficiently large so that the interaction part of the
TGB$_A$ elastic free energy density is dominated by the interaction
of the dislocations in a few nearby grain boundaries. In this
case our formalizm would be reliable.

Provided that that we are in the correct regime, the preceding
section supplies us with all the essential ingredients to compute
the interaction energy of dislocations arranged in the TGB$_A$
structure. To construct an appropriate dislocation source, we need
to combine the sources for individual twist grain boundaries at
positions 0, $\pm l_b$, $\pm 2 l_b$, \ldots\ along the pitch axis.
The grain boundaries at different positions are distinguished by
the direction of defects and, in addition, might be arbitrarily
shifted in the direction perpendicular to the pitch axis.  We may,
through superposition, use the results from the last section to
find:
\begin{eqnarray}\label{fTGBi}
  \fTGBi &=& \frac{D d^2}{2 l_b l_d}
  \Big[\frac{1}{\pi}\sum_{n=1}^\infty
  \BesselK_0\left(\frac{l_d n}{\lambda}\right)\nonumber\\
  &&+ \frac{\lambda}{l_d}\sum_{n=1}^\infty e^{-l_b n/\lambda}\Big].
\end{eqnarray}
We can understand the interaction term by again considering
the confinement energy of the director.  In the case of the full
TGB$_A$ structure, each ``cell'' between the grain boundaries must
be identical.  Thus we may use (\ref{inbetween}) with $B=\theta/2$
to calculate the energy for each cell.  We find
complete agreement with the interaction term above.
\begin{figure}[!hbt]
  \begin{center}
  \epsfig{file=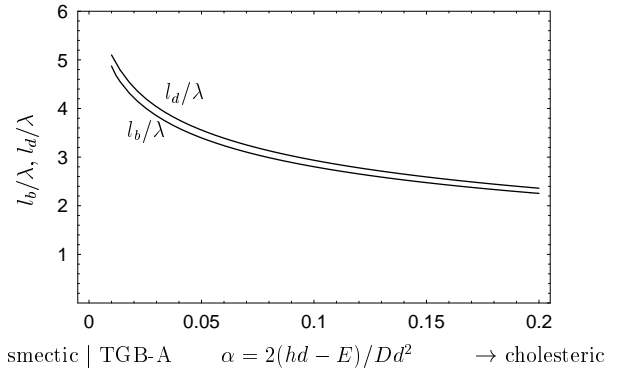,width=80mm}
 \vskip 8truept
  \caption{Dependence of the dislocation spacing within a grain boundary
     $l_d$ and the grain size $l_b$ on the control parameter
     $\alpha = 2 (h d - E)/D d^2$.  At some large value of $\alpha$
there is a transition to the cholesteric phase (at $h_{C2}$) while
$\alpha=0$ corresponds to the smectic--TGB$_A$ transition.}
  \label{graph1}
  \end{center}
\end{figure}
In addition to the elastic energy of interacting screw
dislocations, the total free energy of the \mbox{TGB$_A$}
structure includes two extensive terms that depend only on the
dislocation density $1/(l_b l_d)$ and not on the details of a
particular dislocation arrangement. These terms are the extensive
part of the screw dislocation energy density and the chiral
energy term:
\begin{eqnarray}\label{ftot}
  \ftot &=& \fTGBi + \fdisl + \fch \\ \nonumber
        &=& \frac{D d^2}{2 l_b l_d}
          \left[\frac{1}{\pi}\sum_{n=1}^\infty
          \BesselK_0\left(\frac{l_d n}{\lambda}\right)
          +\frac{\lambda}{l_d}\sum_{n=1}^\infty e^{-l_b n/\lambda}\right]
\nonumber\\ &&+\frac{E}{l_b l_d}-\frac{h d}{l_b l_d},
\end{eqnarray}
where $E$ is the energy cost per unit line of an individual screw
dislocation. In the free energy density (\ref{ftot}), the total
energy cost of dislocations given by the first two terms competes
with the gain in the chiral energy. The twist penetration depth
$\lambda$ sets the length scale for $l_d$ and $l_b$. Inspecting
(\ref{ftot}), we see that the optimal values of $l_d/\lambda$,
$l_b/\lambda$ are controlled by a single combination of the
material parameters

\begin{figure}[!hbt]
  \begin{center}
  \epsfig{file=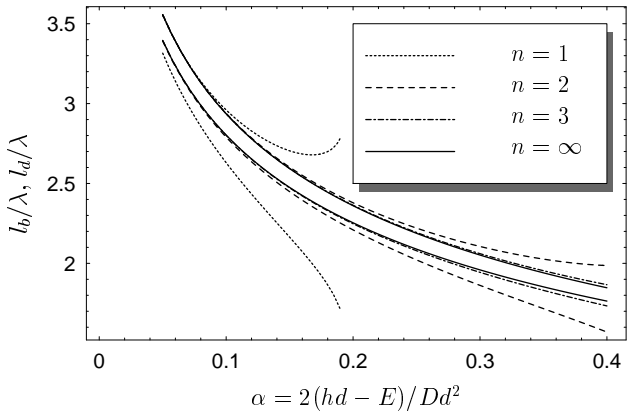,width=80mm}\vskip 8truept
  \caption{Dependence of the minima positions on the number of terms
   kept in the sum over grain boundaries in the free energy density
   (\ref{ftot}). The style of each curve reflects the
   number of terms used in its computation. The upper set of curves
   correspond to
   $l_d/\lambda$, while the lower
   curves correspond to $l_b/\lambda$.
   The solid curves are
   the same curves as in Fig. \ref{graph1}.
   }
  \label{graph2}
  \end{center}
\end{figure}

We minimized of the energy density (\ref{ftot}) with respect to
$l_d/\lambda$ and $l_d/\lambda$ numerically. The results are
presented in Fig. \ref{graph1}.
We can make several observations regarding our results. As is
directly evident from Fig.~\ref{graph1}, there is a range of
$\alpha$ where the preferred values of $l_d$ and $l_b$ are of
several $\lambda$ and, moreover, the ratio $l_d/l_b$ is close to
1. According to the remarks at the beginning of this section,
this kind of geometry validates our computational techniques. To
see how much faraway grain boundaries contribute to the
establishment of the dislocation lattice structure, we computed
the positions of minima for free energy densities obtained from
(\ref{ftot}) by truncating the sums over $n$ to one, two, and
three terms. The results of this computation are presented in
Fig. \ref{graph2}, which demonstrate that in the range $0 <
\alpha < 0.2$, which corresponds to $l_b, l_d > 2.5 \lambda$, the
lattice structure is determined almost entirely by the
interactions of nearest and next-to-nearest twist grain
boundaries.
\begin{equation}
  \alpha = \frac{2 (h d - E)}{D d^2}.
\end{equation}

\begin{figure}[!hbt]
  \begin{center}
  \epsfig{file=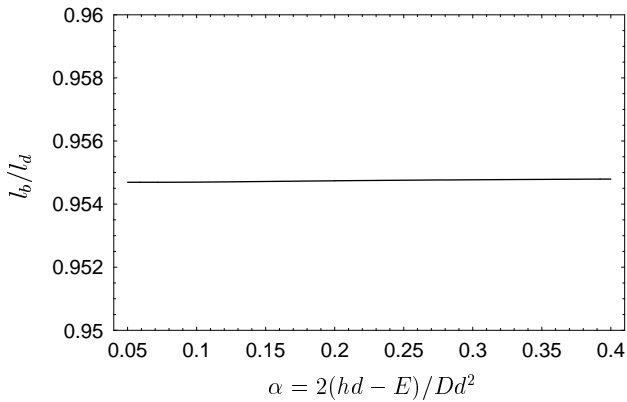,width=80mm}\vskip 8truept
  \caption{Dependence of $l_b/l_d$ on the control parameter $\alpha$.
   Note that it is nearly independent of $\alpha$.}
  \label{graph3}
  \end{center}
\end{figure}

A very interesting result emerges regarding the ratio of the
lattice parameters $l_b/l_d$. It turns out that the ratio is
nearly constant for a very wide range of values of the control
parameter $\alpha$. This prediction should be compared to the
experimental values of $l_b/l_d$ measured by Navailles and
coworkers \cite{Navailles1998}.
\section{Conclusions}
\label{conclusions}
The theory of the screw dislocation interaction in the TGB$_A$
phase in the low-angle grain boundary limit was constructed by
analogy with the theory of vortex interaction in the Abrikosov
phase in the high-$\kappa$ limit. The resulting theory was
applied to the calculation of the lattice parameters of the screw
dislocation arrangement. We found that in this limit the
ratio of the two lattice parameters $l_b/l_d$ remains pratically
constant over a very wide range of the control parameter. The
value of the ratio is 0.95. It is interesting to note this value
is within several thousands of the value obtained by Renn and
Lubensky \cite{Lubensky1988} in the opposite limit $h \to h_{c2}$
for a specific value $K/K_2 = 0$. Whether this is accidental or
not remains to be discovered. In contrast to the Renn and Lubensky
result, the value of $l_d/l_b$ obtained here
is independent of $K_1$ and $K_3$.
This value is consistent with the experimental data of Navailles
and coworkers \cite{Navailles1998}.

\begin{figure}[!hbt]
  \begin{center}
  \epsfig{file=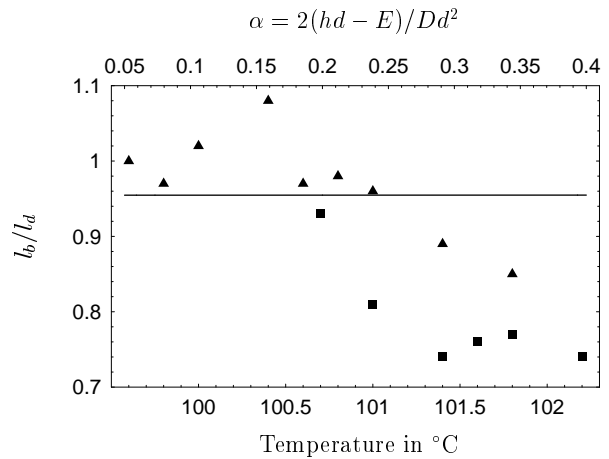,width=80mm}\vskip8truept
  \caption{Experimental dependence of $l_b/l_d$ on temperature.
    The data set marked with triangles is taken from
    \protect\cite{Navailles1998}, the data set marked with boxes was
    provided courtesy of L.~Navailles.
    The first set of data were taken while increasing temperature,
while the latter was taken in a run with decreasing temperature.
    Note that the grain rotation angle increases from $6^\circ$
    for the lowest temperature to $9^\circ$ for the highest temperature.
    }
  \label{graph4}
  \end{center}
\end{figure}

Recent work \cite{Kamien1999} which studied the nonlinear elasticity
of smectic liquid crystals
showed that defects in the same grain boundary had power-law
interactions.  One might expect, in general, that the interaction
between grain boundaries would remain exponential.  It would seem
then that the grain boundaries would move closer together as the
defects in the boundaries would move further apart, casting some
doubt on our result (and on experiment).
In the absence of director modes, the nonlinear
elasticity is \cite{Kamien1999}:
\begin{equation}
 F = \frac{1}{2}\int d^3x\,\left\{B\left[\partial_z u -\frac{1}{2}
 \left(\nabla u\right)^2\right]^2 +
 K\left(\nablap^2u\right)^2\right\}.
\end{equation}
In this nonlinear theory the director and the layer normal are
locked together so that $\vecn = -\nabla \phi/\vert\nabla
\phi\vert$. While a full analysis of the energetics of two
interacting grain boundaries will be the focus of further work,
we can, in the spirit of the analysis at the end of section
\ref{grain}, consider a single, distorted screw dislocation:
\begin{equation}
   u(x,y) = \tan^{-1}\left[\frac{\pi y}{l\tan(\pi x/l)}\right].
\end{equation}
The layer normal of this dislocation relaxes in the usual
way along the $y$-direction but relaxes to its asymptotic value
in the confined region between $x=\pm l/2$.
It is straightforward to calculate the nonlinear energy of this
defect and to find the ``confinement'' energy by subtracting the
$l=\infty$ value.  Expanding in powers of $l^{-1}$ we have
\begin{eqnarray}
  &\displaystyle{\nablap^2 u \approx \frac{\pi^2}{l^2}
 \frac{2xy(x^2-y^2)}{(x^2+y^2)^2}},\\
  &\displaystyle{(\nabla u)^2 \approx \frac{1}{x^2 +y^2}
 + \frac{\pi^2}{l^2}
  \frac{x^2y^2-\textstyle{\frac{2}{3}}x^4}{(x^2+y^2)^2}}.
\end{eqnarray}
It is clear that the $l$-dependence of the confinement energy
scales as $l^{-2}$ which can balance the $l_d^{-2}$ interaction
found between defects in the same grain boundary \cite{Kamien1999}.
We shall explore this further in future work.

\acknowledgments
It is a pleasure to acknowledge fruitful conversations with
L.~Navailles, B.~Pansu, and R. Pindak. IB and RDK were supported
through NSF CAREER Grant DMR97-32963 and NSF Grant INT99-10017.
RDK was supported, in
addition, by the Alfred P. Sloan Foundation and a gift from
L.J.~Bernstein. TCL was supported through NSF Grant DMR97-30405.

\end {document}